\documentclass{tcibook}
\usepackage{fancyhea}
\usepackage{work}
\usepackage{bm}       %    enables bold math symbols  e.g.  \bm{\gamma}
\usepackage{graphicx}
\usepackage{multirow}
\usepackage{lineno}
\usepackage{threeparttable}
\usepackage[normalem]{ulem}
\usepackage{color}
\usepackage[colorlinks = true,
            linkcolor = blue,
            urlcolor  = blue,
            citecolor = blue,
            anchorcolor = blue]{hyperref}

\def\beq{\begin{equation}}
\def\eeq#1{\label{#1}\end{equation}}
\def\eeqn{\end{equation}}

\newenvironment{Eqnarray}%
   {\arraycolsep 0.14em\begin{eqnarray}}{\end{eqnarray}}
\def\beqa{\begin{Eqnarray}}
\def\eeqa#1{\label{#1}\end{Eqnarray}}
\def\eeqan{\end{Eqnarray}}

\let\bar=\overbar

\def\lsim{\mathrel{\raise.3ex\hbox{$<$\kern-.75em\lower1ex\hbox{$\sim$}}}}
\def\gsim{\mathrel{\raise.3ex\hbox{$>$\kern-.75em\lower1ex\hbox{$\sim$}}}}

\def\del{\partial}
\def\Dslash{\not{\hbox{\kern-4pt $D$}}}
\def\dslash{\not{\hbox{\kern-2pt $\del$}}}
\def\pslash{\not{\hbox{\kern-2pt $p$}}}
\def\ETmiss{\not{\hbox{\kern-4pt $E$}}_T}

\def\Dlr{\mathrel{\raise1.5ex\hbox{$\leftrightarrow$\kern-1em\lower1.5ex\hbox{$D$}}}}

\def\MSB{{\bar{M \kern -2pt S}}}
\def\msb{{\bar{\scriptsize M \kern -1pt S}}}

\def\drb{{\bar{\scriptsize D \kern -1pt R}}}

\def\authorlist#1#2{
    \vskip 0.4in
\begin{center}\begin{large} {\bf  #1 } \end{large}
    \vskip 0.2in
              #2
     \vskip 0.2in
   \end{center}
}

\begin{document}

%%  uncomment this line to use line numbers in drafts:
%\linenumbers

\pagenumbering{roman}

\parindent=0pt
\parskip=8pt
\setlength{\evensidemargin}{0pt}
\setlength{\oddsidemargin}{0pt}
\setlength{\marginparsep}{0.0in}
\setlength{\marginparwidth}{0.0in}
\marginparpush=0pt

% The content begins here

\pagenumbering{arabic}

\renewcommand{\chapname}{chap:intro_}
\renewcommand{\chapterdir}{.}
\renewcommand{\arraystretch}{1.25}
\addtolength{\arraycolsep}{-3pt}

\newcommand{\cefgroup}{5}

%% Include the below syntax somewhere in your main .tex file
%% \input{CEF-recommendation-style}
%% (or else otherwise include the commands and environment defined here somewhere)

%% For each recommendation, use the following syntax:
%
% \begin{recs}
%     \rec{A}{B}{C}
% \end{recs}
% 
% A: The number of the recommendation (e.g. 6)
% B: The concise text of the recommendation (e.g. Breese should take a vacation.)
% C: Any supporting text (e.g. Breese has been working very hard for a very long time and has earned a break!)
%
%% If you want to make multiple recommendations within the same pair of thick horizontal bars,
%% Use the \rec command multiple time separated by blank lines
%
% \begin{recs}
%     \rec{A}{B}{C}
%
%     \rec{D}{E}{F}
%
%     \rec{G}{H}{I}
% \end{recs}

\newenvironment{recs}
    {\noindent \begin{minipage}{\textwidth} \rule{\textwidth}{2mm}}
    {\rule{\textwidth}{2mm} \end{minipage}}

\newcommand{\rec}[3]{
\nobreak \noindent \textbf{CEF0\cefgroup~Recommendation #1 -- #2}\\ 
\nobreak \rule{\textwidth}{0.4mm}
%\textit{#3}\\
\nobreak #3
}

\setcounter{chapter}{4} 

%% IMPORTANT:   from this file, refer to the bibliography as   Engagement/CommF05/bibliography.tex   
%%    refer to a figure   A.pdf  as    Engagement/CommF05/figures/A.pdf  .

\chapter{Public Education and Outreach}

\authorlist{S. Demers, K. Jepsen, D. Lincoln, A. Muronga}
   {M. Carneiro, T. Y. Chen}

\section{Executive Summary}
The subject of Public Education and Outreach was included in the Snowmass process for the first time in 2013, covered by the Communication, Education and Outreach topical group. As they mentioned in their report, the group was concerned with efforts to communicate with members of the public, decision-makers, teachers and students~\cite{https://doi.org/10.48550/arxiv.1401.6119}. 

During the 2021 Snowmass process, the Community Engagement Frontier divided discussions of efforts to reach these audiences among three different topical groups. The Public Education and Outreach group focused on the audience of the public. The Public Policy and Government Engagement group focused on the audience of decision-makers. And the Physics Education group focused on the audience of teachers and students. 

In their paper, the 2013 CE\&O topical group recommended: (1) “Making a coherent case for particle physics -- the compelling questions we address, the facilities we need for our research, and the value of particle physics to society”, (2) “Recognizing -- formally and informally -- physicists, postdocs, and students who devote time to CE\&O efforts”, (3) “Developing and increasing access to resources, training activities, and opportunities that engage physicists with policy makers, opinion leaders, the general public, educators and students”, and (4) “Creating a national team dedicated to developing and providing communication and education strategy and resources, [to] supporting and enhancing existing efforts, and to mobilizing a greater fraction of the U.S. community to participate in CE\&O activities.” 

The Public Education and Outreach topical group agrees with these previous recommendations, but laments that they have gone largely unheeded. Seeing how little progress has occurred in this area between this Snowmass process and the last, we focused most intensely on a single issue: what structural changes need to occur, at all levels of the high-energy physics community, to better enable physicists to participate in public engagement. Those structural changes are laid out in detail in our contributed paper, “The need for structural changes to create impactful public engagement in US particle physics.” \cite{https://doi.org/10.48550/arxiv.2203.08916}

The paper gives specific recommendations for changes at the level of research groups, experimental collaborations, conferences, universities and colleges, national laboratories, OSTP, Congress, DOE, NSF, private foundations, AAAS, APS, and DPF. 

\begin{recs}
    \rec{1}{Enact structural changes to better enable public engagement}{In general, we recommend:
\begin{itemize}
\item Providing or financially supporting training in effective public engagement
\item Supporting the creation of public engagement programs that scientists can participate in
\item Codifying the importance of public engagement in official documents such as: 
    \begin{itemize}
    \item Laboratory contracts
    \item Faculty handbooks
    \item Professional society strategic plans
    \item Experimental collaboration constitutions
    \item Merit criteria used by institutions that fund research
\end{itemize}
\item Considering public engagement along with activities such as service and teaching in: 
    \begin{itemize}
    \item Hiring
    \item Tenure
    \item Promotion
    \item Other reviews
    \end{itemize}
\item Funding public engagement work as part of grant proposals 
\item Incorporating public engagement into conferences and meetings in the form of
    \begin{itemize}
    \item Plenary talks
    \item Parallel sessions
    \item Public lectures
    \item Training opportunities for conference participants
    \item Public engagement opportunities for conference participants 
    \end{itemize}
\item Recognizing and rewarding scientists who contribute to public engagement efforts
\end{itemize}
We also recommend that individual scientists encourage others, including peers, mentees and students, by participating in public engagement and discussing its importance.}
\end{recs}

However, structural change was not our only area of focus. Science communication has changed in the years between this Snowmass process and the last. Perhaps the most significant change is a shift from an emphasis on the concept of “public outreach” to an emphasis on the concept of “public engagement,” which the American Association for the Advancement of Science defines as “intentional, meaningful interactions that provide opportunities for mutual learning between scientists and the public.” As the AAAS website explains, “[m]utual learning refers not just to the acquisition of knowledge, but also to increased familiarity with a breadth of perspectives, frames, and worldviews.”\footnote{https://www.aaas.org/focus-areas/public-engagement}

As is explained in the contributed paper “Particle Physics Outreach at Non-traditional Venues,”\cite{https://doi.org/10.48550/arxiv.2203.09585} scientists have innovated in the area of doing more interactive public engagement.
Moving from “outreach” to “engagement” includes making changes such as adding opportunities for conversation between speaker and audience during science talks. But it goes beyond that, to prioritizing efforts to build lasting bonds between a scientific institution and a specific community over efforts to transmit scientific knowledge from one group to the other.

Organizers of a recent International Conference on High Energy Physics, held in Chicago, embodied this philosophy. They began by assessing the needs of local communities, in partnership with people from those communities. They then planned together how ICHEP participants might contribute to addressing those local needs. When community members requested scientists provide science demonstrations, the conference organizers arranged for transportation and equipment, then sent ICHEP participants across the city to talk at schools and libraries about science. The presentations did not necessarily have anything to do with particle physics; the point was to create a way for ICHEP to contribute to the community hosting its meeting. In this report, we advocate for adopting this philosophy of science communication with the public. 

\begin{recs}
    \rec{2}{Update topical group name}{To reflect this shift in priorities, and also to clear up confusion between the goals of our topical group and the topical group focused on physics education, we propose changing the name “Public Education and Outreach” in the next Snowmass process to simply “Public Engagement.”}
\end{recs}

Another big change since the previous Snowmass process involves the way in which scientists think about public audiences. The 2013 CE\&O report categorized audiences mostly by their level of interest in science, defining them as (1) “popular science enthusiasts,” (2) “everyday people,” (3) “parents,” (4) “a geographic audience,” (5) “science skeptics,” and (6) “critics of public funding of science.” 

The Public Education and Outreach group for the 2021 Snowmass process focused more on different kinds of audiences: those made up of members of marginalized communities, who are often excluded, in a variety of ways, from benefiting from scientists’ public engagement.

As indicated by multiple papers contributed to the CEF Diversity and Inclusion topical group, there is a need among scientists to address the lack of diversity in the field of high-energy physics. Specifically designed public engagement can support efforts to recruit and retain scientists from diverse backgrounds. Therefore, our topical group put effort into investigating how scientists can effectively reach members of marginalized communities via public engagement. The Public Education and Outreach topical group wants scientists to know that they have an opportunity to make a real difference, and that if they go into it unprepared and without working in equal partnership with the community they want to reach, they face a real risk of causing lasting harm. 

As explained in the illuminating paper, “‘Not Designed for Us’: How Science Museums and Science Centers Socially Exclude Low-Income, Minority Ethnic Groups,”~\cite{dawson2014not} science communication originally designed for people from a certain cultural background can, in many ways, send the message to people who do not share that cultural background that they are not the intended audience. Unless the communication is redesigned with the new audience in mind, efforts to include that audience may have the opposite effect, reinforcing that audience's sense of exclusion. 

Scientists must include members of a specific community in the planning process for a public engagement effort aimed at that community. And this must be done well, with scientists working toward addressing the community’s needs instead of their own. As described in the Community Engagement Assessment Tool\footnote{https://www.nexuscp.org/wp-content/uploads/2017/05/05-CE-Assessment-Tool.pdf}, in these situations, scientists must move from treating their relationships with communities as “transactional” to treating them as “foundational, continually built between and among people and groups.” To continue to build those relationships, scientists must examine the ways their institutions may contribute to harming the communities they may wish to reach, and must examine whether members of those communities have been excluded from roles, including leadership roles, within their institutions.  
A detailed discussion of the issues can be found in the paper “Building a Culture of Equitable Access and Success for Marginalized Members in Today’s Particle Physics Community”~\cite{MarginalizedCommunities}.

\begin{recs}
    \rec{3}{Reach historically excluded groups}{Physics institutions should consciously plan public engagement to reach historically excluded groups, building long-term relationships with those groups and training physicists to plan activities in partnership with those whom they are hoping to engage.}
\end{recs}

The paper includes the following checklist of questions for scientific institutions to ask when preparing to engage members of marginalized communities:

Consider the audience:
\begin{itemize}
    \item Who specifically are we hoping to reach with this event? Why are we hoping to reach these communities?
    \item How can we plan this event to make it maximally beneficial to these communities? What elements of this plan can we continue to use in other events?
    \item What are the best ways to communicate about this event with members of these communities? Can we continue going to those same channels to communicate about other events?
    \item Have we created a process by which we take time to evaluate the success of the event after it concludes? 
    \item What metrics (both qualitative and quantitative) will we use? Which of these metrics will we continue to use in evaluating other events?
\end{itemize}
    
Identify and remove barriers:
\begin{itemize}
    \item Are there logistical barriers (e.g. time of day, day of the week, public transportation access, affordability, safety concerns, financial barriers) to our events that make them inaccessible to these communities? What will we do to address these barriers?
    \item Have we allowed adequate lead time and budget to make this event accessible to all members of these communities, including those with disabilities? Have we identified partnership or staffing needs required to make the event accessible?
\end{itemize}
    
Value partnerships:
\begin{itemize}
    \item What members of these communities will make good partners in this event? Have we made sure they’re involved in planning the event? Have we secured an adequate budget to support fair compensation for our partners as co-creators of the event, prior to requesting their labor?
    \item Do any members of these communities work for our institution? If they do, do they work in roles with decision-making power (e.g. managerial positions), or do they work primarily in service roles? If members of these communities do not work at our institution, or work only in lower-level positions, is our institution making any effort to change this?
    \item Are members of these communities who work for our institution participating in this event? If so, are they receiving the support they need to take on this effort and fulfill their other job duties? Do they have decision-making power over the planning and execution of the event? Are they being fairly compensated and recognized for their efforts?
\end{itemize}
    
Build lasting relationships:
\begin{itemize}
    \item Is this event a part of a larger effort to build relationships with members of these communities? If so, what is the long-term plan? Who will be responsible for enacting it?
    \item Are there ways in which our institution is causing harm to members of these communities? If so, how is our organization working to change this?
    \item How are representatives of our institution involved in these communities outside of this event? Are there ways our institution can work with members of these communities on their priorities, even ones that do not directly benefit our institution?
\end{itemize}

We request the organizers of Snowmass focus in the coming years on finding ways to implement these recommendations for needed structural change, and on supporting efforts to build lasting relationships between scientific institutions and members of marginalized groups. We recommend that the American Physical Society’s Division of Particles and Fields take up the mantle of monitoring progress on these goals in the years between this Snowmass process and the next.

We hope that the rest of this topical group report, which details our other explorations during the Snowmass process, will be helpful to the co-conveners of the Public Engagement topical group during the next Snowmass process.

\section{Summary of activities}

The 2021 Snowmass process started in 2020. During our first sessions, the four co-conveners of the Public Education and Outreach topical group decided on two overall goals:
\begin{itemize}
\item To have public education and outreach in particle physics recognized as important scientific activities and supported at all levels
\item To make public education and outreach a part of every practicing particle physicist’s job description \end{itemize}
During the Snowmass process, we conducted a survey to gather information about community priorities and issues; we wrote and received letters of interest; we considered perspectives from several invited guests; and we wrote one contributed paper, co-wrote another contributed paper, and received a third contributed paper. 

\subsection{Survey}

At the end of 2020, the Public Education and Outreach topical group conducted a survey of members of the physics community. We received 358 responses from people who identified themselves as students, postdocs, non-tenured and tenured physicists, and other staff. 

The majority of respondents said they had participated in some kind of outreach or engagement in the last two years. The most common activities mentioned were public lectures, science communication on social media, institutional communication, K-12 education, media interviews, and public tours. Some people wrote that they had participated in citizen science and in organizing public engagement activities. 

When asked about the factors that prevent them from doing outreach and engagement, respondents overwhelmingly said they did not have time. The second most common factor respondents chose was that doing outreach and engagement provided no career benefit. After that, people said that they lacked access to events, lacked training, and lacked support. A few respondents wrote that they were uncomfortable with public speaking, sometimes due to a language barrier. 

When asked about their goals for their outreach and engagement, respondents said that they wanted to reach underserved groups and wanted to demonstrate openness to the public. Several respondents wrote in that they wanted to share their enthusiasm, explain the scientific method to the public, or inspire the next generation of scientists.

The two tactics respondents were most likely to have used in their outreach and engagement were storytelling and talking about their own motivations for pursuing physics. 

Multiple respondents offered advice to scientists interested in getting involved in outreach and engagement. They recommended starting small and building up experience with public speaking. They recommended finding and plugging into established programs and organizations, even ones not focused on physics or STEM. 

\subsection{Letters  of Interest (LOI)}

The Public Education and Outreach topical group was tagged in 13 LOIs:

“Snowmass Early Career Longterm Organization”: This LOI, submitted by the Early Career topical group, discussed the formation of a long-term organization of the EC HEP community. We invited the group to meet with us, as we wanted to encourage them to adopt public education and outreach as a priority in their organization. The LOI was withdrawn to be included in the EC topical group discussions.

“Science / Society: considering new paradigms of planning for public engagement and communication” and “Science outreach and the underrepresented public”: These LOIs addressed the intersection of public engagement and diversity \& inclusion. The Public Education and Outreach and Diversity \& Inclusion topical groups explored these concepts in the contributed paper “Building a Culture of Equitable Access and Success for Marginalized Members in Today’s Particle Physics Community”\cite{MarginalizedCommunities}.

“The Cosmic Ray Extremely Distributed Observatory as a new quality public engagement and edutainment environment”, “CREDO-Maze: Multi-stage Global Network of School EAS Mini-arrays ('the quest for the unexpected')”, and “An extensible, experiment-agnostic file format to facilitate educational access to HEP datasets”: Some of these LOIs were consolidated and presented as contributed papers in the Physics Education topical group~\cite{https://doi.org/10.48550/arxiv.2203.08809, https://doi.org/10.48550/arxiv.2203.08919}.

“Progress on High School Physics Outreach”: This was found to be more aligned with the goals of the Physics Education topical group~\cite{https://doi.org/10.48550/arxiv.2203.10953}.

“The African School of Fundamental Physics and Applications” and “Expanding Fermilab’s international outreach through European networks”: These LOIs focused on public education and outreach in countries outside the United States. They were consolidated into a contributed paper, ``The Necessity of International Particle Physics Opportunities for American Education'' \cite{https://doi.org/10.48550/arXiv.2203.09336}, that was submitted via the Physics Education topical group.

“Public Education and Outreach” and “The CERN-IARI Project and New Opportunities for Integrated Arts Research Collaborations at Universities and National Laboratories”: We invited the authors of these LOIs on festivals, music/arts and physics to present to our topical group. The ideas in these papers appeared in a contributed paper, “Particle Physics Outreach at Non-traditional Venues”\cite{https://doi.org/10.48550/arxiv.2203.09585}.

Our topical group wrote two LOIs: “Structural changes for public engagement with particle physics and particle physics communication” and “Ensuring the conditions that encourage effective participation in public engagement.” Ideas from these LOIs appeared in the contributed paper “The need for structural changes to create impactful public engagement in US particle physics”\cite{https://doi.org/10.48550/arxiv.2203.08916}.

\subsection{Perspectives}

During the Snowmass process, the Public Education and Outreach topical group met with several individuals, as shown in Table~\ref{CEF-CommF05-table:table1}, to solicit a wide range of perspectives on how particle physics outreach and engagement with the community could be improved:

\begin{table}[!h]
\centering
\begin{tabular}{l|l|l}
\hline\hline
\textbf{Date}   &  \textbf{Group} & \textbf{Individuals} \\ \hline
24-Nov-2020 & Cosmic Rays & Mark Adams, QuarkNet \\
& & Lukasz Bibrzycki, CREDO \\
& & Mike Mulhearn, CRAYFIS \\
& & Ivan Sidelnik, LAGO \\
& & Justin Vandenbroucke, DECO \\
& & Tadeusz Wibig, CREDO-Maze \\ \hline
20-Jul-2021 & Art & Emily Coates, Dancer \\
& & Bill Collins, Music $+$ Climate. \\
& & Larry Lee, Musician \\
& & Lindsay Olsen, Textile Artist \\
& & Georgia Schwender, FNAL Art Director \\ \hline
5-Oct-2021 & APS & Jim Gates \\ \hline
23-Nov-2021 & APS & Jonathan Bagger \\ \hline
12-Dec-2021 & NSF & Mark Cole \\ \hline
18-Jan-2022 & DOE & Michael Cooke, Alan Stone, Crystal Yeh \\ \hline
22-Feb-2022 & DOE & Rick Borchelt, Allison Eckhart \\
\hline\hline
\end{tabular}
\caption{Meetings to discuss improvements in physics outreach and engagements.}
\label{CEF-CommF05-table:table1}
\end{table}

For each topic, the speaker was encouraged to present any prepared remarks, after which they fielded questions from the committee. In the text below, we have tried to distill the most salient insights.

\textbf{Cosmic Rays}: We spoke with the PIs of an assortment of programs attempting to bring cosmic rays studies to the public and classrooms. Each program had a different focus, but all of the programs were aimed at either students (with a guiding teacher) or members of the science-interested public. These programs could be emulated by other groups doing, for example, gravitational-wave research. We encouraged the PIs from different programs to organize a cosmic-ray outreach community. 

\textbf{Art}: We invited to our meeting a group of artists and an art director, all of whom have been involved with art-and-science partnerships at institutions. They recommended getting to know one’s audience prior to organizing public presentations. They recommended making presentations interactive via Q\&A sessions with both artists and scientists.

\textbf{Jim Gates}: Physicist Jim Gates was at the time of the meeting president of the American Physical Society and has been involved in public education and outreach. Symmetry, an institutional publication about particle physics and astrophysics, covered much of his presentation to our group in an article.\footnote{https://www.symmetrymagazine.org/article/jim-gates-gives-back}

\textbf{Jonathan Bagger}: Physicist Jonathan Bagger was at the time of the meeting CEO of the American Physical Society. He provided guidance on disseminating messages about public education and outreach outside of the Snowmass process, including submitting a letter to the editor to APS News and presenting to the annual APS leadership conference.

\textbf{Mark Coles}: Physicist Mark Coles was at the time of the meeting NSF HEP program manager. He advised that within the NSF, there was a strong bias for public outreach efforts focused on diversity. This reflected the priorities of the current administration. 

Coles noted that NSF education programs are administered and evaluated by people with a formal background in K-12 and undergraduate education. These individuals have a different set of criteria, values, and terminology than those common in the particle physics environment.  Coles advised that if individuals from HEP were interested in accessing funding from educational programs, it would be helpful to form collaborations with education departments within individual universities.  

\textbf{Michael Cooke}: Physicist Michael Cooke was at the time of the meeting a DOE program officer. He noted that education and outreach are not part of the Department of Energy mandate. As such, any proposals aimed solely at engaging the public are unlikely to be supported by DOE. However, one important DOE goal is to broaden and diversify the scientific workforce. Outreach programs with a strong diversity component are received more favorably, whether aimed at the public or students in an educational environment. The preference at DOE is to support education programs in a formal environment and at a more advanced level–for example, supporting visiting faculty from underrepresented institutions, internships for underrepresented undergraduates, or fellowships for similar graduate students.  

Cooke noted that, while DOE was unlikely to provide support for public engagement at individual universities, it does indirectly support public engagement at national laboratories via overhead funds, used at the discretion of individual laboratory management. 

\textbf{Rick Borchelt}: Rick Borchelt was at the time of the meeting the director of Communications and Public Affairs at DOE Office of Science. He described the formation of the Science Public Engagement Partnership (SciPEP) with the Kavli Foundation. This effort is focused on conducting research on how to communicate with the public about basic research. 

\subsection{Contributed papers}

The Public Education and Outreach topical group wrote one contributed paper, co-wrote a second contributed paper, and received a third contributed paper from an outside group.

\subsubsection{“The need for structural changes to create impactful public engagement in US particle physics”}

The Public Education and Outreach topical group focused its attention on writing the contributed paper “The need for structural changes to create impactful public engagement in US particle physics”~\cite{https://doi.org/10.48550/arxiv.2203.08916} The paper explores six types of institutions that assert significant influence in the physics ecosystem:

\begin{itemize}
    \item the particle physics community—including research groups, experimental collaborations, and scientific conferences; 
    \item universities and colleges—including departments, schools and colleges, and leadership; 
    \item the national laboratories; 
    \item governmental institutions, including the Office of Science and Technology Policy and Congress; 
    \item institutions that fund research, including the National Science Foundation, the Department of Energy, and private foundations; 
    \item professional bodies and societies, including the American Association for the Advancement of Science, the American Physical Society, and the APS Division of Particles and Fields.
\end{itemize}

It details why each of these institutions benefits or could benefit from public engagement by physicists. And it explains why each of these institutions is influential in determining whether physicists participate in public engagement. 

As mentioned in the executive summary, the paper lists individual recommendations for each of these institutions to better enable physicists to participate in public engagement. 

\subsubsection{“Building a Culture of Equitable Access and Success for Marginalized Members in Today’s Particle Physics Community”}

Given the interest that members of the HEP community expressed in reaching underserved groups with their public engagement efforts, the Public Education and Outreach topical group also collaborated with the Diversity \& Inclusion topical group to address public engagement in a contributed paper about marginalized communities~\cite{MarginalizedCommunities}.
 
The paper made the point that, unless scientists make a concerted effort to reach members of marginalized communities with their public engagement, this work will serve to reinforce the status quo of unequal access to HEP among members of the public and limited diversity within HEP itself.
 
Engaging the public and effectively reaching members of marginalized communities are skills that must be learned. The public engagement section of the paper advocates for scientists to learn about the difference between outreach—which is more of a one-way transfer of information that primarily values the knowledge of the scientist—and engagement—which is more of a two-way dialogue that values the knowledge of both the scientist and the audience, and which can be an effective method for building relationships and trust.
 
The paper also recommends scientists learn about cultural competence, focusing on developing a deep understanding of the factors that affect whether people from a certain cultural background benefit from efforts to engage the public in STEM. The paper recommends scientists read “‘Not Designed for Us’: How Science Museums and Science Centers Socially Exclude Low-Income, Minority Ethnic Groups”~\cite{dawson2014not} to better understand this issue.
 
Developing the cultural competence needed to design effective public engagement requires studying how different barriers intersect and interact to discourage or prevent people from participating in public engagement activities. It requires rejecting an attitude of promotion and adopting one of collaboration with these communities, striving to pursue community-driven goals and build lasting relationships.

\subsubsection{“Particle Physics Outreach at Non-Traditional Venues”}

Jim Cochran, John Huth, Roger Jones, Paul Laycock, Claire Lee, Lawrence Lee, Connie Potter, and Gordon Watts contributed a third paper to the Public Education and Outreach topical group~\cite{https://doi.org/10.48550/arxiv.2203.09585}. The paper introduces “The Big Bang  Collective,”\footnote{https://www.lancaster.ac.uk/physics/outreach/big-bang-collective/} started in 2016, which brings “Physics Pavillions” to “many of the best known music and culture festivals across Europe and the UK.” The paper points out that this model of public engagement brings particle physics to where people are and can reach an audience that might not otherwise engage with science outreach.

The paper outlines the importance of taking into account the audience demographics and festival style when planning activities. Past Physics Pavillions have included:

\begin{itemize}
    \item curated programs of talks; 
    \item dedicated workshops, such as “build your own cloud chamber”;
    \item live virtual link-ups to organizations such as CERN and NASA;
    \item outdoor hands-on science activities throughout the day;
    \item virtual reality tours of LHC experiments.
\end{itemize}

The paper advocates for an extension of this model to the United States, where we have many music and cultural festivals and a large number of publicly engaged scientists. The paper notes a number of festivals that could serve as pilots, including Bonnaroo Music and Arts Festival in Tennessee and the Northwest Folklife Festival and Bumbershoot Festival in Seattle. 

%%%%%%%%%%%%%%%%%%%%%%%%%%%%%%%%%%%%%%%%%%

%  If you would like to use BibTEX for the bibliography, please feel free to do so.  It is not required.

%  To use BibTeX,

%    1.  uncomment the following two lines,
%    2.  comment out everything below from  \begin{thebibliography}{99}   to \end{thebibliography).
%    3.  create the file  myreferences.bib in this directory, and process this file in the usual way

\bibliographystyle{JHEP}
\bibliography{Engagement/CommF05/myreferences} 

%%%%%%%%%%%%%%%%%%%%%%%%%%%%%%%%%%%%%%%%%

%\begin{thebibliography}{99}

%\input Engagement/CommF05/bibliography.tex

%\end{thebibliography}

% \input Engagement/CommF06/Policy.tex
% \input Engagement/CommF07/Society.tex

%%%%%%%%%%%%%%%%%%%%%%%%%%%%%%%%%%%%%%%%%%%%%%%%%%

\end{document}